# Password Authentication Scheme with Secured Login Interface

**A.T. Akinwale and F.T. Ibharalu**
Department of Computer Science, University of Agriculture,
Abeokuta, Nigeria
atakinwale@yahoo.com

**ABSTRACT.** This paper presents a novel solution to the age long problem of password security at input level. In our solution, each of the various characters from which a password could be composed is encoded with a random single digit integer and presented to the user via an input interface form. A legitimate user entering his password only needs to carefully study the sequence of code that describe his password, and then enter these code in place of his actual password characters. This approach does not require the input code to be hidden from anyone or converted to placeholder characters for security reasons. Our solution engine regenerates new code for each character each time the carriage return key is struck, producing a hardened password that is convincingly more secure than conventional password entry system against both online and offline attackers. Using empirical data and a prototype implementation of our scheme, we give evidence that our approach is viable in practice, in terms of ease of use, improved security, and performance.
**KEYWORDS:** Alphanumeric, Password authentication, Algorithm and Combination

## Introduction

A password is a secret word or phrase that gives a user access to computer resources such as programs, files, messages, printers, internet, etc. The password assists in ensuring that unauthorized user do not access a restricted resource. Ideally, the password should be something that nobody could guess. In practice, most people choose a password that is easy to remember

77

such as their names or their initials. This is one reason it is relative easy to break into most computer systems [Has05].

The main drawback in the design of many password mechanisms arises from the fact that password lengths are usually small or short. This makes it easy to spy and memorize passwords through the monitoring of computer keystrokes physically or through eavesdropping. For example, in a university environment each student may be assigned a password to protect his resources within the school computer system. Using the password during lecture time may endanger the security of his password because of the presence of other students around him. Similarly, login page of Yahoo! requires each user to enter his/her password online. Such 'online password' can also be compromised by a hacker who monitors online keystroke sequence programmatically.

In a big corporation, each personnel accesses computer resources through password. If the password is very short, as this is usually the practice, the secretary to the manager may spy her boss password and thereby have access to her boss corporate resources.

Personal identification number (PIN) is always used by various banks to allow their customers access to their online bank accounts. For the customers' convenience, PINS are often short and in many cases only numeric of up to eight digits (ATM uses only four numeric digits). This is why many banks issue warnings to their customers to study the automated teller machine (ATM) surroundings properly before approaching it for use to protect their PINS. The implication of this is that PINs are not safe since they can be subject to attack (spying) by third parties.

The advantage of short length characters for PINs and password is that it is easy to be remembered by the user. This advantage is a problem since it makes it easier for attacker to memorize short keystroke of PIN and password. In this paper, an algorithm is developed to provide a strong security support for both short and long character-password at input level.

## 1. The Structure of the System

The structure of the secured password authentication scheme, as depicts in figure 1, consists of four parts, namely – the input characters, algorithms, legitimate users password database and form interface,.

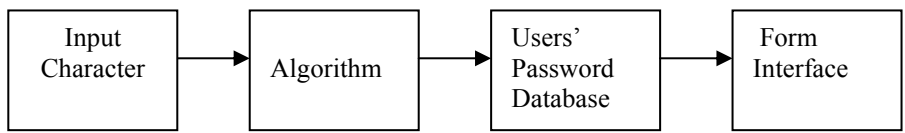

Figure 1: Structure of secured password authentication scheme

## 1.1. Input Characters

The system accepts all printable ASCII characters, which may consist of lower and upper case (A-Z, a-z), numeric digits (0-9), and special characters (+ - _ ^ , # % etc).

## 1.2. The Algorithms

### 1.2.1. Definitions

Let the user input code be denoted by $I_n = c_1c_2c_3…c_n$, and its length be defined as $n = L(c_1c_2\,c_3…c_n$. Let **X** be the list of all the password characters i.e. A,…,Z, a,…,z, 0-9, symbols} and **Y** the set of the corresponding randomly generated numbers such that each value of **X** is assigned a random value from **Y**.

### 1.2.2. Constraint

The set of randomly generated numbers $S = \{r : 0 \leq r \leq 9, r \in \mathbf{Z}\}$ is such that $n(X)$ is evenly divisible by $n(S)$, i.e. $\dfrac{n(X)}{n(S)} = \mathbf{d}$ must be an integer. Also, the frequency of each $y_i = \mathbf{d}\ \forall i$

### 1.2.3. Algorithm for generating all characters corresponding to each of the user input code

Computation of $G(c_i)$, the set of characters corresponding to each $c_i$ of user input where $c_i \in I_n$ and i =1, 2,…,n.

79

Input: User input code, $c_1, c_2, c_3, \ldots, c_n$
Method: Let $n = L(c_1 c_2 c_3 \ldots c_n)$
Compute $G(c_i) = \{List(y_i)\}, i = 1, 2, 3, \ldots, n$

where $List(y_i) = \begin{cases} x_i, & \text{if } c_i = y_i, \\ null, & \text{otherwise.} \end{cases}$ and $List = \bigcup x_i \to y_i$

Output: The set of password characters corresponding to each $c_i$.

### 1.2.4. Algorithm for password authentication

Definition:
Password database D[p] is a sorted list of p legitimate users-passwords that are allowed access to the system.
Input: D[p]
  $n = L(c_1 c_2 c_3 \ldots c_n)$ i. e. user password length
  $G(c_i)$, the set of password characters of each input code
Method:

Dim index[n] as integer
Switch (k)
Begin
  case 1:
        For index [0] = 0 to $L(G(c_0)) - 1$;
            tempPassword = strncpy (tempPassword, $G(c_0) + index[0]$, 1);
            if (search (D, tempPassword)) == true)
            begin
               login (); exit Password Authentication;
            end
         next $index[0]$;
       write "password not found"
  case 2:
For index [0] = 0 to $L(G(c_0)) - 1$;
        For index [1] = 0 to $L(G(c_1)) - 1$;
            tempPassword = strncpy (tempPassword,
                $G(c_0) + index[0]$, 1)
                    + strncpy (tempPassword, $G(c_1) + index[1]$, 1)

80

```
                        if (search (D, tempPassword)) == true)
                        begin
                            login (); exit Password Authentication;
                        end
                next index[1];
    next index[0];
        write "password not found"
```

• • •

case n:
For index [0] = 0 to $L(G(c_0))-1$;
    For index [1] = 0 to $L(G(c_1))-1$;
        For index [2] = 0 to $L(G(c_2))-1$;
        • • •
            For index [n-1] = 0 to $L(G(c_{n-1}))-1$;
                tempPassword =
                strncpy (tempPassword, $G(c_0)+index[0]$, 1)
            + strncpy (tempPassword, $G(c_1)+index[1]$, 1)
            + strncpy (tempPassword, $G(c_2)+index[2]$, 1)
                • • •
            + strncpy (tempPassword, $G(c_{n-1})+index[n-1]$, 1)
            if (search (D, tempPassword)) == true)
            begin
                exit Password Authentication;
                login ();
            end
        next index[n-1];
        • • •
        next index[2];
    next index[1];
  next index[0];
write "password not found"

## 2. Database

The database, **D**, contains all the passwords of valid users. The database has been designed to accept passwords of length up to 255 characters without spaces but the implementer must restrict size of users' passwords to a reasonable length, for example twelve characters, for easy computation.

## 3. Form interface

Visual Basic 6.0 (VB) was used for the implementation of the algorithm. From the algorithm, a form interface was generated as shown in figure 2. The form shows the password characters (in black) with their corresponding numeric code (in red).

Figure 2: Form interface

The interface contains alphanumeric (A-Z, a-z, 0-9) and special characters (+, -, *, /, etc) each of which is labeled with a single numeric digit code between 0 and 9. The single digit labels are generated randomly and are equally distributed.

An user must study the interface form carefully, and then enter the numeric digits corresponding to his actual password characters. Upon completion of the input the user hits the enter key. This triggers the algorithm to regenerate a new input form followed by authentication of the last input password. The input interface form regeneration is necessary to harden the password entry system and make it extremely difficult, if not impossible, for attackers to spy.

## 4. Analysis of the scheme

Assuming an user password is 'Lagos(2006)', and that the current input interface form generated by our algorithm is as shown in figure 2, then the sequence of digits corresponding to this password that the user must enter is 27318081174. Looking at figure 2, there would be many characters that would correspond to each of the numeric code. However, the algorithm has been constrained (see 2.2.2) to distribute the numeric digits (0, 1, 2… 9) equally amongst the permissible password characters.

Table 1 gives all the characters corresponding to each of the user input code 27318081174. Algorithm 2.2.4 computes each password combination (LBVTAHATTBI, LBVTAHATTBP, …, ^/wp_(_pp/! ) derivable from table 1 and validates it with the password database. If the current combination is validated successfully, the process terminates and the user is logged in, otherwise the next password combination is tried for validation. If all combinations are tried with failures, then the user is denied access.

The total password combination from the scheme is $d^n$ which could be a very large value for a carefully planned password system. This will make it extremely difficult for any attacker to spy physically or eavesdrop electronically because the user keystrokes are not directly related to the actual password.

| User password (unknown) | L | a | g | O | s | ( | 2 | 0 | 0 | 6 | ) |
|---|---|---|---|---|---|---|---|---|---|---|---|
| User input code (as fig. 2) | 2 | 7 | 3 | 1 | 8 | 0 | 8 | 1 | 1 | 7 | 4 |
| Characters corresponding to each of the user input code | L | B | V | T | A | H | A | T | T | B | I |
| | Y | C | 4 | W | F | M | F | W | W | C | P |
| | H | E | 7 | O | 2 | N | 2 | O | O | E | Z |
| | T | 6 | 9 | 1 | 3 | e | 3 | 1 | 1 | 6 | + |
| | { | 8 | c | B | n | u | n | b | b | 8 | ) |
| | } | a | g | D | s | z | s | d | d | a | % |
| | - | v | i | O | y | [ | y | o | o | v | = |
| | ^ | / | w | P | _ | ( | _ | p | p | / | ! |

Table 1: Characters corresponding to the user input code **27318081174**

## Conclusion

Randomly generated digits corresponding to password characters on input interface form that makes it impossible for user password to be spied at input level have been proposed.

The scheme evenly distributes randomly generated numeric digits among the password characters so that it can produce many possible password combinations that are difficult for attackers to spy or electronically eavesdrop. Integrating the modules into existing systems where password authentications are required will improve the security and integrity of the password systems.